\documentclass[%
10pt,
 oneocolumn,
 nofootinbib,
 amsmath,amssymb,
 aps,
floatfix,
unsortedaddress
]{revtex4}

\usepackage{graphicx,color}
\usepackage{subfig}
\usepackage{dcolumn}
\usepackage{bm}

\usepackage[pdfpagelabels, pdfencoding=auto, psdextra]{hyperref}
\hypersetup{%
 pdfsubject=Paper,
 pdfkeywords={nuclear physics} {Pionless EFT} {Two-Nucleon} {Lattice} {L\"usher formula},
 unicode = true,
 breaklinks = true,
 colorlinks = true,
 linkcolor = blue,
 citecolor = blue,
 menucolor = blue,
 citecolor = blue,
 urlcolor = blue
}

\newcommand{\nc}{\newcommand}
\nc{\bp}{{\bf p}}
\nc{\bpp}{{\bf p}'}
\nc{\bx}{{\bf x}}
\nc{\bn}{{\bf n}}
\nc{\Nn}{N_{\bf n}}
\nc{\NTn}{N^T_\bn}
\nc{\Ndn}{N^{\dag}_\bn}
\nc{\Delp}{\mathcal{P}_L}
\nc{\Delpp}{\Delp'} 
\nc{\bpL}{{\bf p}_L}
\nc{\bppL}{{\bf p}'_L}
\nc{\Cz}{C_0^L}
\nc{\Cd}{C_2^L}
\nc{\ANC}{\text{ANC}}
\nc{\at}{a^{(^3 S_1)}}
\nc{\rt}{r^{(^3 S_1)}}

\graphicspath{{./}}
\begin{document}

\title{A Novel Regularization Scheme for Nucleon-Nucleon Lattice Simulations with Effective Field Theory}

\author{M. Ahmadi}
 \email{Masoumehahmadi@ut.ac.ir}
\affiliation{
 Department of Physics, University of Tehran, PO Box 14395-547, Tehran, Iran
}%

\author{M. R. Hadizadeh}
 \email{mhadizadeh@centralstate.edu}
\affiliation{
College of Engineering, Science, Technology and Agriculture, Central State University, Wilberforce, OH, 45384, USA
}%
\affiliation{Department of Physics and Astronomy, Ohio University, Athens, OH, 45701, USA}

\author{M. Radin}
 \email{Radin@kntu.ac.ir}
\affiliation{
 Department of Physics, K. N. Toosi University of Technology, PO Box 16315-1618 Tehran, Iran
}%

\author{S. Bayegan}
\email{Bayegan@ut.ac.ir} \thanks{(Corresponding author)}.
\affiliation{
Department of Physics, University of Tehran, PO Box 14395-547, Tehran, Iran
 }%

\date{\today}%

\begin{abstract}
We propose a new regularization scheme to study the bound state of two-nucleon systems in Lattice Effective Field Theory. Inspired by continuum EFT calculation, we study an exponential regulator acting on the leading-order (LO) and next-to-leading order (NLO)
 interactions, consisting of local contact terms.
By fitting the low-energy coefficients (LECs) to deuteron binding energy and the asymptotic normalization coefficient (ANC) on a lattice simulation, we extract the effective range expansion (ERE) parameters in the $^3S_1$ channel to order $p^2$. We explore the impact of different powers of the regulator on the extracted ERE parameters for the lattice spacing $a=1.97$ fm. 
Moreover, we investigate how the implementation of the regularization scheme improves the predicted ERE parameters on the lattice spacing in the range of $1.4 \le a \le 2.6$ fm. 
Our numerical analysis indicates that for lattice spacing greater than $2$ fm, the predicted observables are very close to the experimental data.
\end{abstract}

\maketitle

\section{Introduction}\label{introduction}
Nuclear lattice effective field theory (NLEFT) is a model-independent and precision controlled approach for the calculation of bound and scattering state properties in nuclear physics \cite{lahde2019nuclear}. The novel combination of lattice methods with an effective field theory approach has been pursued successfully for few- and many-body systems.

The first attempts for an exact solution of infinite nuclear matter using Monte Carlo methods are performed in Ref. \cite{muller2000nuclear}, indicating that energy and saturation properties of symmetric nuclear matter can be reproduced from lattice simulations. 
The ab initio techniques combine the Monte Carlo methods with the low-energy EFT, known as chiral effective field theory. Based on these approaches, our information for the scattering of light nuclei, and the ground-state properties of light-, medium-mass nuclei, as well as neutron matter has been compromised \cite{lu2019essential, borasoy2007lattice, borasoy2008chiral, lee2009lattice, epelbaum2009modern, machleidt2011chiral}.  
To improve the efficiency of large-scale calculations of nucleus-nucleus scattering and reactions using Monte Carlo calculations, the adiabatic projection method is developed on lattice \cite{pine2013adiabatic, elhatisari2016nucleon}.
The accuracy and efficiency of the method are tested on fermion-dimer scattering calculations in lattice EFT.

The bound state of two nucleons on a lattice, mainly in the $S-$wave channel, is formulated in pionless EFT at the NLO \cite{harada2016numerical}. The lattice spacing dependence of the RG flows is studied while the deuteron binding energy and the ANC are being fixed. 
L\"uscher has shown how one can connect the quantities obtained on a finite volume to the infinite volume physical observables by connecting the box size dependence of energy eigenvalues on a lattice to the effective range parameter and the scattering length ~\cite{luscher1986volume}.
The exact solution of L\"uscher formula for the energy-levels of the two-nucleon system on a lattice with periodic boundary conditions for the extraction of scattering parameters has been implemented by Beane {\it et al.} in a pionless EFT approach~\cite{beane2004two}. They have shown that lattice simulations with $L \ge 15$ fm will provide information on the scattering lengths and effective ranges straightforwardly. Whereas the extraction of data from lattice simulations with $L \le 10$ fm requires direct matching to $p \cdot cot \delta_0$ in the spin-singlet channel and considering the mixing between the $S-$ and $D-$wave remains challenging.
The impact of the topological finite-volume corrections in lattice calculations of three-nucleon bound state \cite{bour2011topological}, the elastic scattering of fermion-dimer \cite{bour2012benchmark}, and also neutron-deuteron scattering at the very low energies ~\cite{ rokash2013finite} are studied in a pionless EFT approach. 

One of the main challenges of lattice calculations is the necessity to eliminate errors caused by the non-vanishing lattice spacing. 
One approach to eliminate the lattice artifacts is including the irrelevant higher-dimensional operators into the lattice action, which leads to faster convergence to the continuum limit \cite{klein2018lattice}.
Since the lattice spacing serves as a natural UV regulator for the theory, another practical strategy is the application of a regulator to utilize the smearing of the contact interactions.
Klein {\it et al.} have shown that the application of different regularization schemes leads to the lattice spacing independence of observables for a wide range of the lattice spacing in the range $0.5 \le a \le 2.0$ fm ~\cite{klein2015regularization}. This study is performed at the leading order of pionless and pionfull EFT. The extension of the calculations to the two-, three-, and four-body sectors to study the lattice spacing dependence up to next-to-next-to-leading order (N$^2$LO), including two- and three-nucleon interactions, is performed in Ref. \cite{klein2018tjon}. The binding energy correlation of triton and helium-4 is studied for various lattice spacings $a=1.97, 1.64, 1.32$ fm, and it is shown how the convergence towards the Tjon line is reached for smaller lattice spacing.
A systematic study of neutron-proton scattering, in terms of the computationally efficient radial Hamiltonian method, is studied on a lattice EFT up to N$^2$LO \cite{alarcon2017neutron}. A regularization scheme is applied only to the LO contact interactions. 
The lattice spacing dependence of the scattering observables is explored for lattice spacings ranging from $a = 1.97$ fm down to $a = 0.98$ fm, and it is shown at $a = 0.98$ fm, the lattice artifacts appear to be small.
In a recent study by Eliyahu {\it et al.}, the effect of the finite lattice size on the binding energies of light nuclei is explored by the construction of pionless EFT at the LO, where a gaussian regulator is applied on the contact terms \cite{eliyahu2019extrapolating}.

In this paper, we propose a regularization scheme, inspired by continuum EFT calculations, to study the two-nucleon systems on a lattice and extract the ERE parameters for a wide range of lattice spacing.
In Sec. \ref{2N_Formalism}, we briefly review the formalism of two-nucleon bound state on a lattice, projected in the $^3S_1$ channel, using pionless EFT up to NLO.
By introducing the Lagrangian and Hamiltonian of the two-nucleon system on a lattice, the explicit form of the Lippmann-Schwinger equation is presented by considering the contact interactions between nucleons.
In Sec. \ref{Luscher}, the procedure of extraction of physical ERE parameters from finite volume energy eigenvalues is discussed. 
Our numerical results for the lattice energy eigenvalues obtained for different lattice spacing parameters and different numbers of lattice nodes are presented in Sec. \ref{Results}. Moreover, a new regularization scheme is introduced, and the impact of the regularization scheme on the ERE parameters is studied in detail.
A conclusion is provided in Sec. \ref{conclusion}. All the energy eigenvalues obtained for different lattice spacing parameters are provided in the Appendix \ref{Appendix_Energies}.

\section{Two-Nucleon in $^3 S_1 $ channel in pionless Lattice EFT up to NLO} \label{2N_Formalism}
At very low energies where the nucleon momentum is much smaller than the pion mass, i.e., $Q \ll m_\pi$, few-nucleon systems are not sensitive to the details of the nucleon-nucleon interactions. So, an EFT is constructed by low energy degrees of freedom and the Lagrangian is formulated as all contact interactions between nucleons that are allowed by symmetry.
In this section, we consider the NLO Lagrangian of pionless EFT. The nucleon-nucleon interactions are defined by an infinite number of local operators with an increasing number of derivatives acting on the nucleon fields. The isospin SU(2) symmetric and nonrelativistic Lagrangian in the continuum is given by
\begin{eqnarray} \label{Lagrangian}
\mathcal{L} &=&
N^{\dag}\left[i\partial_{t}+\dfrac{{\nabla}^2 }{2M}\right]N \cr
&-& C_0 \left(N^T P^k N)^{\dag}(N^T P^k N\right)  \cr
&+& C_2 \left[(N^T P^k N)^{\dag}(N^T P^k \overleftrightarrow{\nabla}^2 N)+h.c.\right],
\end{eqnarray}
where $N$ denotes the nonrelativistic nucleon field, $M$ is nucleon mass, the low-energy constants (LECs) $C_0$ and $C_2$ are the zero-range interaction strengths, and $\overleftrightarrow{\nabla}^2 =\overleftarrow{\nabla}\cdot\overleftarrow{\nabla}-
2\cdot\overleftarrow{\nabla}\cdot\overrightarrow{\nabla}+\overrightarrow{\nabla}\cdot\overrightarrow{\nabla}$.
$P^k = \frac{1}{\sqrt{8}} \sigma_2 \sigma^k \tau_2 $, with the vector indices $k=1,2,3$, is the projection operator for $^3 S_1 $ channel, where $\sigma_2$ and $\tau_2$ are the Pauli matrices acting on the spin and isospin spaces, respectively.
The Hamiltonian corresponding to the Lagrangian of Eq. (\ref{Lagrangian}) is given by
\begin{eqnarray}\label{Hamiltonian}
H &=& \int  d^3 x \bigg[ N^{\dag} \left (\frac{{-\nabla}^2 }{2M} \right) N \cr
&& \hspace{1.1cm}  + C_0 \left(N^T P^k N)^{\dag}(N^T P^k N\right)  \cr
&&\hspace{1.1cm}  -C_2 \left ( (N^T P^k N)^{\dag}(N^T P^k \overleftrightarrow{\nabla}^2N)+h.c.\right) \bigg].
\end{eqnarray}
To study the bound state of two-nucleon systems on a lattice, we utilize a cubic box of side length $L$ with periodic boundary conditions. The lattice spacing between lattice nodes is $a$, so that $L=N_{s}a$, where $N_s$ is the number of nodes in each spatial direction.
As it is shown in Ref. \cite{harada2016numerical}, in order to transform the Hamiltonian of Eq. (\ref{Hamiltonian}) from the continuum to a dimensionless Hamiltonian on a lattice, one needs to apply the following substitutions
\begin{eqnarray}
&& N(\bx)\rightarrow \Nn  a^{-3/2},
\quad
\bx \rightarrow  \bn a,
\quad
\int d^3 x\rightarrow a^3 \sum_\bn  ,
\quad \cr
&&H \rightarrow H_L a^{-1},
\quad
M\rightarrow M_L a^{-1},
\quad C_0 \rightarrow \Cz a^2,
\quad
C_2 \rightarrow C^L_2 a^4 ,
\end{eqnarray}
where $\bn  = (n_1,n_2, n_3)$ is a three-dimensional vector with integer components and
$\Cz$, $C^L_2$ and $M_L$ are dimensionless parameters, corresponding to parameters $C_0$, $C_2$ and $M$ in continuum.
The lattice Hamiltonian $H_L $ can be obtained in terms of dimensionless quantities as
\begin{eqnarray} \label{Hamiltonian_lattice}
H_L &=& \sum_\bn  \bigg[ \dfrac{-1}{2M_L }\Ndn   \nabla^2_L \Nn    \cr
&& \hspace{.9cm} + \Cz \left(\NTn  P^k \Nn )^{\dag}(\NTn  P^k \Nn  \right)  \cr
&& \hspace{.9cm} - C^L_2 \left\{(\NTn  P^k \Nn )^{\dag}(\NTn  P^k \overleftrightarrow{\nabla}_L^2\Nn )+h.c.\right\}\bigg],
\end{eqnarray}
where $\nabla^2_L$ and $\overleftrightarrow{\nabla}_L^2$ represent the discretization of the dimensionless Laplacian. By considering the nucleon operator $\Nn$ in momentum space as
 \begin{eqnarray}
\Nn  =\frac{1}{N_s^{3/2}}\sum_{\bpL } e^{i\bpL  \cdot \bn }a_{\bpL} ,
\end{eqnarray}
the lattice Hamiltonian of Eq. (\ref{Hamiltonian_lattice}) leads to
\begin{eqnarray} \label{lattice_Hamiltonian}
H_L &=& \sum_{\bpL}\frac{\Delp}{2M_L }a^{\dag}_{\bpL}a_{\bpL} \cr
&+& \frac{1}{N_s^3 } \sum_{\bpL ,\bppL}  \left ( \Cz +4C_2 ^{L}
(\Delp+\Delpp) \right)
 \left ( a^{\dag}_{\bpL}P^k a^{\dag}_{-\bpL} \right)
 \left (a_{\bppL} P^k a_{-\bppL}  \right).
\end{eqnarray}
The momentum argument $\Delp $ obtained from the free nucleon lattice action, improvement up to ${\cal O} (a^4)$, defined as \cite{lee2009lattice}
\begin{eqnarray} \label{P^2_equation}
\Delp \equiv 2\sum_{i=1}^3 \biggl ( \omega -\omega_1 \cos(p_i) + \omega_2 \cos(2 p_i)-\omega_3 \cos (3 p_i)\biggr).
\end{eqnarray}
where the components of the lattice momentum $\bpL  \equiv (p_1, p_2, p_3)$ under the periodic boundary condition takes the values
\begin{eqnarray}
p_i = \dfrac{2\pi}{N_s} \hat{p}_i,  \quad - \dfrac{N_s}{2}  < \hat{p}_i  \leq \dfrac{N_s}{2}, \quad i=1,2,3.
\end{eqnarray}
As it is shown in Ref. \cite{lee2009lattice}, the hopping coefficients $\omega_i$ in the improved free nucleon action eliminate lattice artifacts in the Taylor expansion of single-nucleon dispersion relation around $\bpL=0$ up to the indicated order.
The coefficients $\omega_i$ for different level of improvement up to ${\cal O} (a^4)$, are listed in Table. \ref{Table_hopping_coefficients}.
It should be noticed that the ${\cal O} (a^{2n})$-improved lattice action corresponds to a lattice derivative which contains $2n+2$ nearest neighbors, or a total of $2n+3$ lattice sites. It means unimproved, ${\cal O} (a^2)$-improved, and ${\cal O} (a^4)$-improved actions are corresponding to three-, five-, and seven-point formula, respectively.
By considering the lattice Hamiltonian of Eq. (\ref{lattice_Hamiltonian}), the lattice form of Lippmann-Schwinger equation for two-nucleon bound state can be obtained as \cite{harada2016numerical}
\begin{eqnarray}\label{LS_lattice}
\psi(\bpL)= \frac{1}{E_L - \dfrac{\Delp }{M_L} } \cdot \dfrac{1}{N_s^3 }
\sum_{\bppL} \left ( \Cz +4C_2 ^L(\Delp +\Delpp) \right) \psi(\bppL),
\end{eqnarray}
where $E_{L}=Ea$ is the dimensionless two-nucleon binding energy and $\psi(\bpL)$ is the discretized two-nucleon wave function.

 \begin{table}[htp!]
  \caption{Hopping coefficients $\omega_i$ for different levels of improvement up to ${\cal O} (a^4)$ in the free nucleon lattice action \cite{lee2009lattice}.}
\centering
\begin{ruledtabular}
\begin{tabular}{lccccccccccc}
 && unimproved &&& ${\cal O} (a^2)$-improved &&& ${\cal O} (a^4)$-improved \\
\hline
$\omega$ && $1$ &&& $5/4$ &&& $49/36$ \\
$\omega_1$ && $1$ &&& $4/3$ &&& $3/2$ \\
$\omega_2$ && $0$ &&& $1/12$ &&& $3/20$ \\
$\omega_3$ && $0$ &&& $0$ &&& $1/90$ \\
 \end{tabular}
\end{ruledtabular}
 \label{Table_hopping_coefficients}
 \end{table}

\section{Extraction of effective range expansion parameters in lattice} \label{Luscher}
By solving the discretized form of the Lippmann–Schwinger equation of (\ref{LS_lattice}), one can obtain the two-nucleon energy eigenvalues on the lattice.
In the following, we briefly show how the L\"uscher formula can be used to extract the ERE parameters in $^3 S_1$ channel by having the deuteron binding energy spectrum on the lattice.
L\"uscher has shown how one can connect the physical quantities in a finite volume to the real physics by connecting the box size dependence of the energy eigenvalues in a finite volume to the infinite volume scattering matrix.
As it is shown in Ref. \cite{beane2004two},
the low-momentum behavior of the $S-$wave phase shift $\delta_0$, for two-nucleons with a relative momentum $p$, can be described by the following ERE
\begin{eqnarray} \label{ERE}
p \cdot  \cot \delta_0 (p) &=& -\dfrac{1}{\at} + \dfrac{1}{2} \rt \ p^2 + \dots \cr
&=& \dfrac{1}{\pi L}S(\eta),
\end{eqnarray}
where $\at$ and $\rt$ refer to the scattering length and the effective range, respectively. $S(\eta) $ is the three-dimensional zeta function with the dimensionless argument $\eta= \left(\dfrac{L p}{2\pi} \right)^2$.
For $| \eta |  < 1$, $S(\eta) $ can be expanded in powers of $\eta$ as
 \begin{eqnarray} \label{S_function}
S(\eta) = -\dfrac{1}{\eta} + S_0 + S_1 \eta+S_2 \eta^2 +S_3 \eta^3 + \dots
 \end{eqnarray}
where the first few coefficients $S_i$ are given as
\begin{eqnarray}
&&S_0 =-8.913631, \quad S_1 =16.532288, \quad S_2 =8.401924, \quad S_3 =6.945808, \cr
&&S_{4}= 6.426119,  \quad S_{5}=6.202149, \quad S_{6}= 6.098184, \quad
S_{7}=6.048263 .
\end{eqnarray}
By considering the connection between the two-nucleon energy levels $E_2 =E_L /a  =  \frac{p^2 }{M}$ and the argument $\eta$, {\it i.e.}, $E_L =  \frac{\eta a}{M}  (\dfrac{2\pi}{L})^2$, one can obtain a set of $\eta$ for a set of energy eigenvalues $E_L$ obtained for a given lattice parameter $a$ and different values of $N_s$ or the box side length $L$. By using Eq. (\ref{S_function}), the function $\dfrac{1}{\pi L}S(\eta)$ can be obtained for different values of $\eta$ dictated by energy eigenvalues $E_L$. Finally by using a linear fitting to Eq. (\ref{ERE}), one can extract the ERE parameters $\at$ and $\rt$.

\section{Numerical Results}\label{Results}

\subsection{LECs and different levels of improvement in the lattice momentum argument $\Delp$}

In this section, we study the effect of different levels of improvement, up to ${\cal O}(a^4)$, in the lattice momentum defined in Eq. (\ref{P^2_equation}) to solve the lattice form of Lippmann-Schwinger Eq. (\ref{LS_lattice}). To this aim, we solve the discretized Lippmann-Schwinger equation for the lattice spacing $a=1.97$ fm with the number of nodes $N_s=20$.
The equation can be solved with both direct and Lanczos methods. Our numerical analysis shows that the runtime of the calculations with the direct approach increases exponentially with the number of nodes $N_s$. For instance, a direct diagonalization of the kernel of Eq. (\ref{LS_lattice}) for $N_s=20$ takes about $90$ minutes, while an iterative solution with the Lanczos technique (see Appendix C2 of Ref. \cite{hadizadeh2012binding}) takes about $1$ second, both performed on a single-node CPU desktop.
While we are convinced that both methods yield the same results for lattice deuteron binding energy and wave function, we perform all the calculations with the Lanczos technique to save runtime.
The Eq. (\ref{LS_lattice}) is an eigenvalue equation in the form of $\lambda \ \psi = {\cal K}(E_L) \cdot \psi$ with the eigenvalue $\lambda = 1$. Since the kernel of the equation ${\cal K}(E_L)$ is energy dependent, the solution of the eigenvalue Eq. (\ref{LS_lattice}) can be started by an initial guess for the energy $E_L$ and the search in the binding energy is stopped when $| 1 - \lambda | \le 10^{-6}$.

The LEC $\Cz$ at LO is fitted to deuteron binding energy $E_d = -2.224575$ MeV, while at NLO, both LECs $\Cz$ and $\Cd$ are determined simultaneously by fitting to deuteron binding energy as well as the asymptotic normalization coefficient $\ANC = 0.249424$ fm$^{-0.5}$. The value of ANC is extracted from the expression for the $S-$wave asymptotic normalization coefficient $\ANC = \frac{1}{\sqrt{4 \pi}} \sqrt{ \frac{2k_0}{1-\rt k_0} }$ \footnote{The factor $\frac{1}{\sqrt{4 \pi}}$ comes from the normalization of the spherical harmonics.} \cite{phillips2000improving}, with $k_0 = \sqrt{M |E_d|}$ and the experimental value of $\rt = 1.759(5)$ fm.
Similar to the procedure performed in Ref. \cite{harada2016numerical}, the ANC parameter can been extracted by fitting the numerical lattice deuteron wave function $\psi(\bpL)$ to the analytical wave function $\psi(\bpL) = A + \frac{B}{ M_L |E_L| + \Delp}$, with $\ANC = \frac{B}{4\pi}$.
To extract the physical values of LECs $\Cz$ and $\Cd$, Eq. (\ref{LS_lattice}) is solved for a wide range of coefficients $\Cz$ and $\Cd$. In Table \ref{LECs_improvement}, we have listed the obtained LECs at LO and NLO for different levels of improvement.
As we can see at LO, the improvements up to ${\cal O}(a^2)$ and ${\cal O}(a^4)$ lead to about $17\%$ and $23\%$ increasing in the absolute value of $\Cz$, respectively.
While at NLO, the improvements up to ${\cal O}(a^2)$ and ${\cal O}(a^4)$ lead to about $4\% ~(14\%)$ and $6\% ~ (18\%)$ increasing (decreasing) in the absolute value of $\Cz$ ($\Cd$), respectively.
In order to minimize the lattice artifacts in our numerical study, for the rest of the paper we use ${\cal O}(a^4)$-improvement in the lattice momentum $\Delp$.
\begin{table}[htp!]
\caption{The LECs $\Cz$ and $\Cd$ obtained at LO and NLO for different levels of improvement in the lattice momentum $\Delp$, defined in Eq. (\ref{P^2_equation}), to reproduce deuteron binding energy $E_d = -2.224575$ MeV and $\ANC =0.249424$ ~fm$ ^{-0.5}$ for the lattice parameter $a = 1.97$~fm and $N_s=20$.}
\centering
\begin{ruledtabular}
\begin{tabular}{ccccccccccccc}
 Improvement Level & $ \Cz$ & $ \Cd  \cdot 10^{-2}$ & $ E_2$ (MeV) & \ANC ~(fm$ ^{-0.5}$)  \\
\hline
  & \multicolumn{4}{c}{{\bf LO }}    \\ \cline{2-5}
unimproved & $-0.49656112$ & $ 0$ & $-2.224575$ &  $ 0.186319$    \\
${\cal O}(a^2)$-improved & $-0.5792460$ & $ 0$ & $-2.224574$ &  $ 0.200483$    \\
${\cal O}(a^4)$-improved & $-0.60920561$ & $ 0$ & $-2.224575$ &  $ 0.204060$    \\
\hline
  & \multicolumn{4}{c}{{\bf NLO }}    \\ \cline{2-5}
  unimproved & $-1.49576$ & $+3.711484$ & $-2.224574$ &  $ 0.249423$    \\
${\cal O}(a^2)$-improved & $-1.56064$ & $+3.173400$ & $-2.224575$ &  $ 0.249425$    \\
${\cal O}(a^4)$-improved & $-1.57907$ & $+3.0453214$ & $-2.224575$ &   $ 0.249425$   \\
\end{tabular}
\end{ruledtabular}
\label{LECs_improvement}
\end{table}

\subsection{A New Regularization Scheme in Lattice}\label{Regularization}
\begin{figure*}[htp!]
\centering
  \includegraphics[width=.6\textwidth]{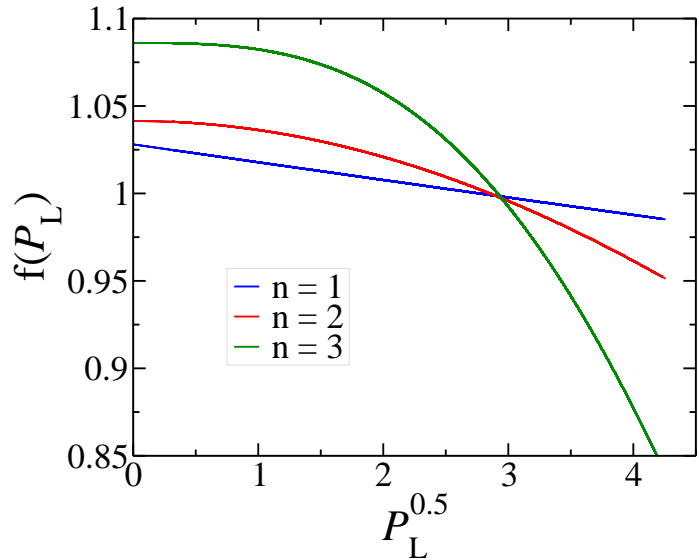}
\caption{The functional form of the regulator $f(\Delp)$, defined in Eq. (\ref{regulator}), for $n=1, 2, 3$ and the regulator parameter $ b=0.01$. }
\label{Fig.regulators}
\end{figure*}

In this section, we introduce a new regularization scheme and study its impact on the ERE parameters $\at$ and $\rt$ obtained from the lattice energy eigenvalues $E_L$ for different values of lattice spacing.
Inspired by continuum EFT calculations \cite{epelbaum2000nuclear}, we consider the exponential regulators in the lattice nucleon-nucleon interactions $V^L_{NN} (\Delp,\Delpp)$ as
 \begin{equation}\label{regularization}
V^L_{NN} (\Delp,\Delpp) \to V^L_{NN} (\Delp,\Delpp)  \cdot f(\Delp) \cdot f(\Delpp),
\end{equation}
where the regulators are defined as
\begin{equation} \label{regulator}
 f(\Delp) = \dfrac{1}{f_0}  \exp(-b\cdot \Delp^{n/2} /n); \quad f_0 = \dfrac{1}{N_s^3} \sum_{\bpL}  \exp(-b\cdot \Delp^{n/2} /n) .
\end{equation}
It should be noticed that the $\Delp^{n/2}$ is calculated from the lattice momentum argument $\Delp$, defined in Eq. (\ref{P^2_equation}).
The regulator parameter $b$ is dependent on the lattice spacing parameter $a$ and is defined as $b \cdot a^3 = {\cal A}$.
A typical value of the regularization parameter in our calculations for the lattice spacing $a=1.97$ fm is $b=0.01$, which leads to the constant parameter ${\cal A} =  7.645373 \cdot 10^{-2}$ fm$^3$.
In Fig. \ref{Fig.regulators}, we have shown the regulator $f(\Delp)$ as a function of the lattice momentum
$\Delp^{0.5}$ for three exponential powers $n=1, 2, 3$ with the regulator parameter $b=0.01$. The lattice momentum argument $\Delp$ is obtained for $N_s=20$.
To study the effect of the regulators on the prediction of the ERE parameters $\at$ and $\rt$, we solve Eq. (\ref{LS_lattice}) with different regulator powers for the lattice spacing $a=1.97$ fm and $N_s=20$. For each power of the regulator, we refit the LECs in such a way that $\Cz$ and $\Cd$ reproduce the deuteron binding energy and the ANC. Then by having the LECs, we resolve Eq. (\ref{LS_lattice}) to calculate the energy eigenvalues $E_L$ for smaller values of $N_s$, in the domain $4 \le N_s\le 20$. Finally, by applying the L\"uscher formula, as discussed in Sec. \ref{Luscher}, we extract the ERE parameters from the energy eigenvalues.
We implement the same steps at the LO, where the only LEC parameter $\Cz$ reproduces the deuteron binding energy, and we have no control over the ANC.
In Table \ref{Table.1.97-regulators}, we have presented our numerical results for the prediction of the ERE parameters $\at$ and $\rt$, with different powers of the regulator. At the NLO, deuteron binding energy and ANC are both used as inputs to extract the LECs $\Cz$ and $\Cd$, while at the LO, the only input to extract $\Cz$ is deuteron binding energy.
As we can see, applying the regulator leads to a correction in the ERE parameters, and it seems the power $n=1$ leads to more corrections than $n=2$ and $n=3$.
\begin{table}[htp!]
\caption{Deuteron binding energy, ANC, and the ERE parameters $\at$ and $\rt$ calculated for the lattice spacing parameter $a =  1.97$ fm. $n,b$ indicates the parameters of the regulator, defined in Eq. (\ref{regulator}). The numbers in parentheses are the uncertainties in the last digits.}
\centering
\begin{ruledtabular}
\begin{tabular}{ccccccccccccccc}
Order & $ n, b $ & $ \Cz$ & $ \Cd  \cdot 10^{-2}$ & $ E_2$ (MeV) & \ANC~(fm$ ^{-0.5}$) & $\at$ (fm) &
$\rt$ (fm)   \\
\hline
 LO & $ 1, 0$ & $-0.60920561$ & $ 0$ & $-2.224575$ &  $ 0.204060$  & $4.577(7)$ & $0.496(8)$   \\
 LO & $ 1, 0.01$ & $-0.6017484$ & $ 0$ & $-2.224573$ &  $ 0.199551$  & $4.652(7)$ & $0.621(7)$   \\
 LO & $ 2, 0.01$ & $-0.5929415$ & $ 0$ & $-2.224576$ &  $ 0.195223$  & $4.624(7)$ & $0.580(8)$   \\
 LO & $ 3, 0.01$ & $-0.569575$ & $ 0$ & $-2.224575$ &  $ 0.184159$  & $4.64(1)$  & $0.60(1)$   \\
\hline
 NLO & $ 1, 0$ & $-1.57907$ & $+3.0453214$ & $-2.224575$ &   $ 0.249425$  & $5.35(3)$ & $1.65(2)$  \\
  NLO & $ 1, 0.01$ & $-1.59677$ & $+3.2940765$ & $-2.224575$ & $ 0.249424$ & $5.43(5)$ & $1.74(4)$   \\
  NLO & $ 2, 0.01$ & $-1.587607$ & $+3.4237993$ & $-2.224574$ & $ 0.249423$ & $5.41(4)$ & $1.73(3)$   \\
 NLO & $ 3, 0.01$ & $-1.574284$ & $+3.945544$ & $-2.224575$ & $ 0.249427$ & $5.42(3)$ & $1.76(2)$   \\
\hline
Experiment & $-$ & $-$ & $-$ & $-2.224575$ & $ 0.249424$ & $ 5.424(4)$ & $ 1.759(5)$ \\
\end{tabular}
\end{ruledtabular}
\label{Table.1.97-regulators}
\end{table}

\begin{figure*}[htp!]
    \centering
    \subfloat{{\includegraphics[width=0.45\textwidth]{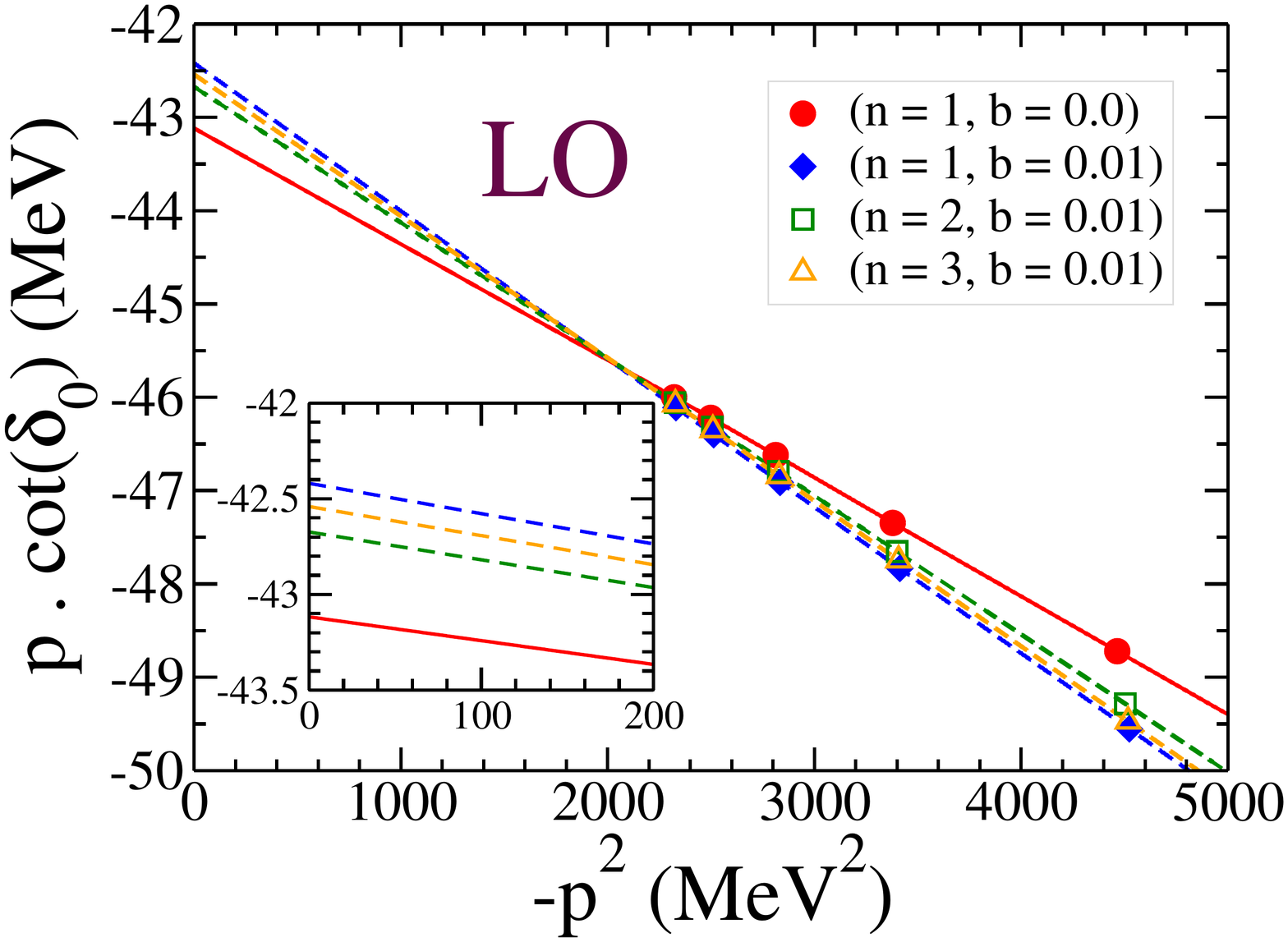} }}%
    \subfloat{{\includegraphics[width=0.45\textwidth]{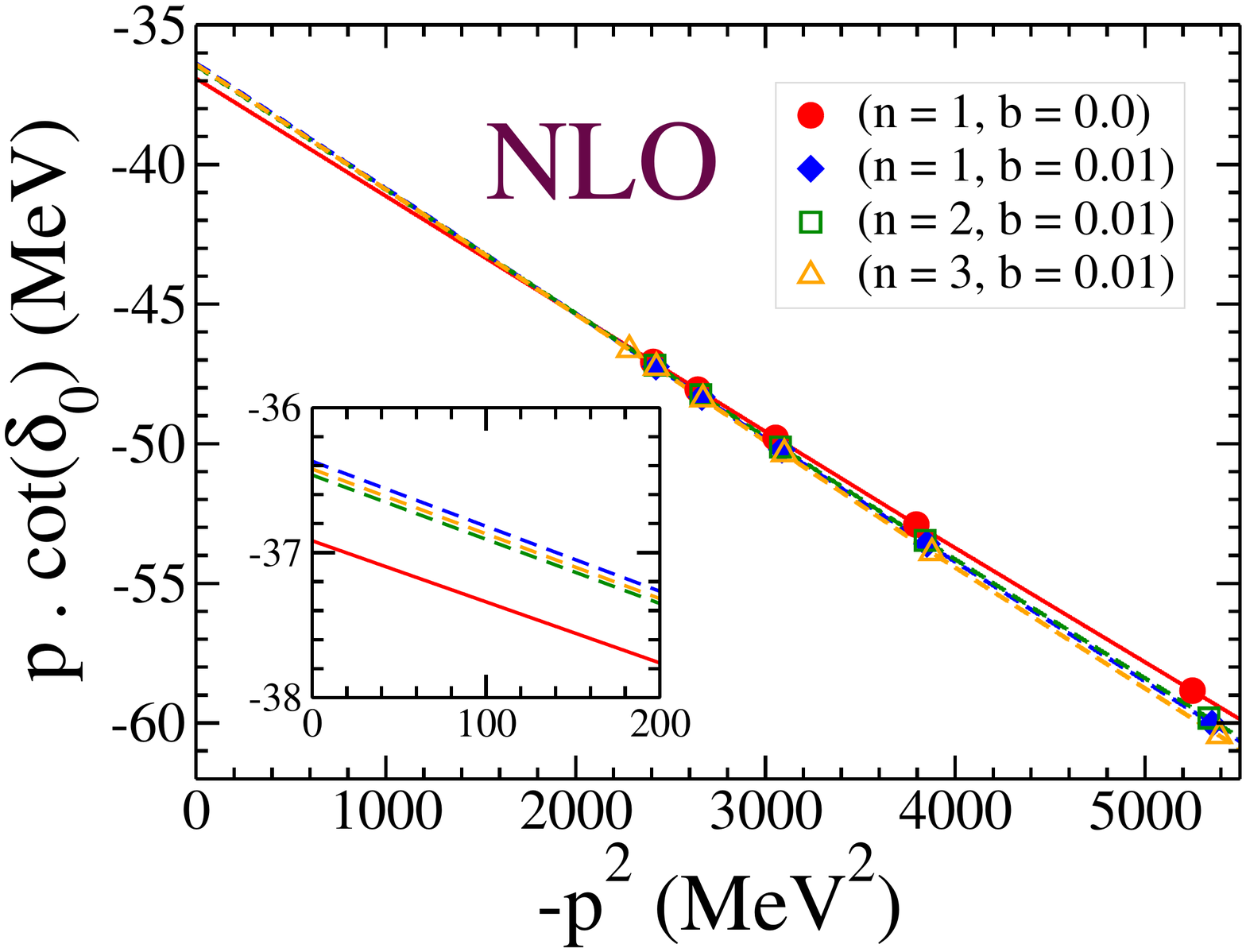} }}%
    \caption{Effective range function in the $^3S_1$ neutron-proton channel for the lattice spacing $a=1.97$ fm, with and without regulators. The solid red line indicates the results obtained by bare contact interactions, while the blue, green, and orange dashed lines are corresponding to the results obtained with regularized interactions with powers $n=1$, $2$, and $3$, respectively.}
    \label{fig_ERE_1.97}
\end{figure*}

\begin{figure*}[htp!]
    \centering
    \subfloat{{\includegraphics[width=0.45\textwidth]{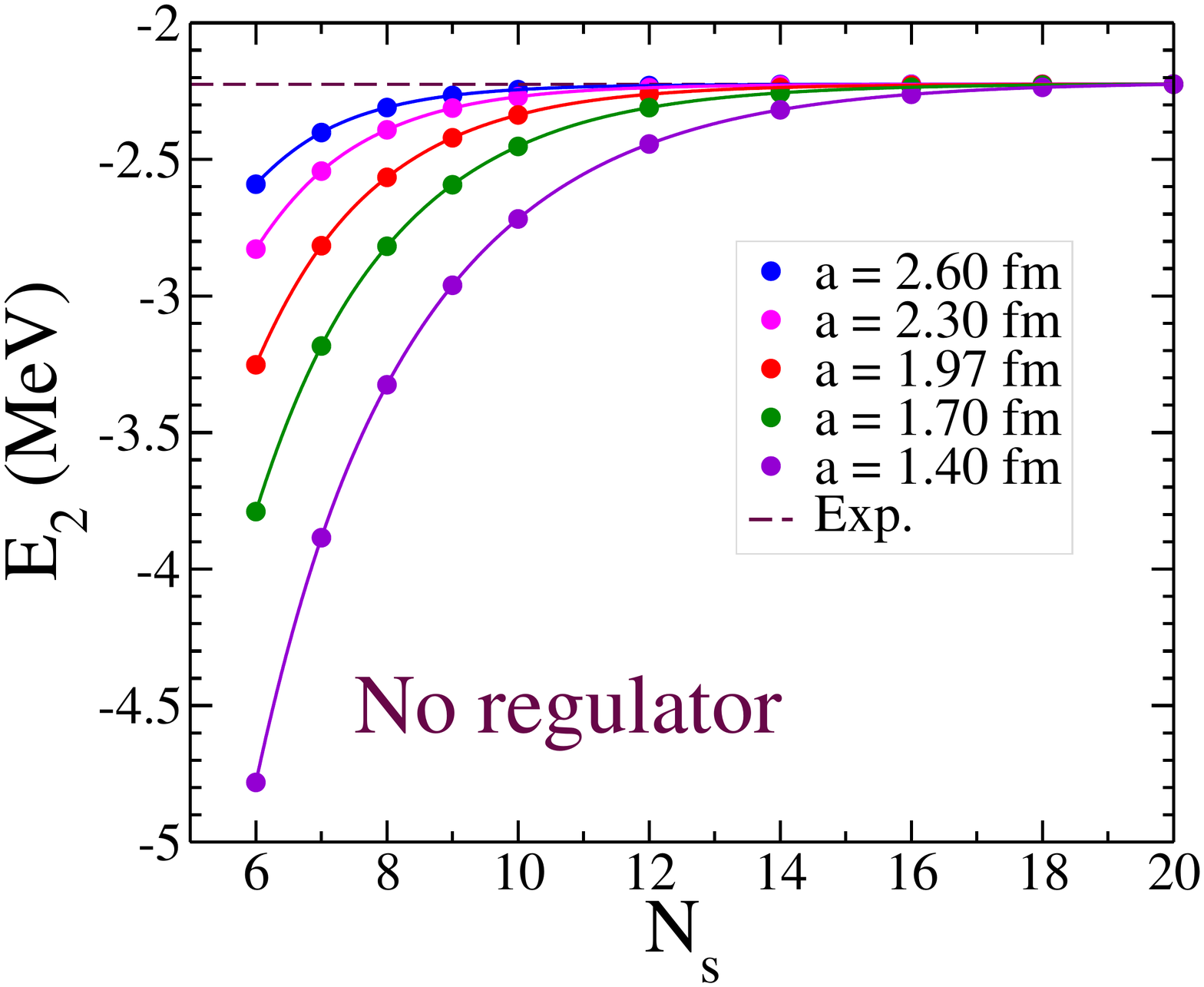} }}%
    \subfloat{{\includegraphics[width=0.45\textwidth]{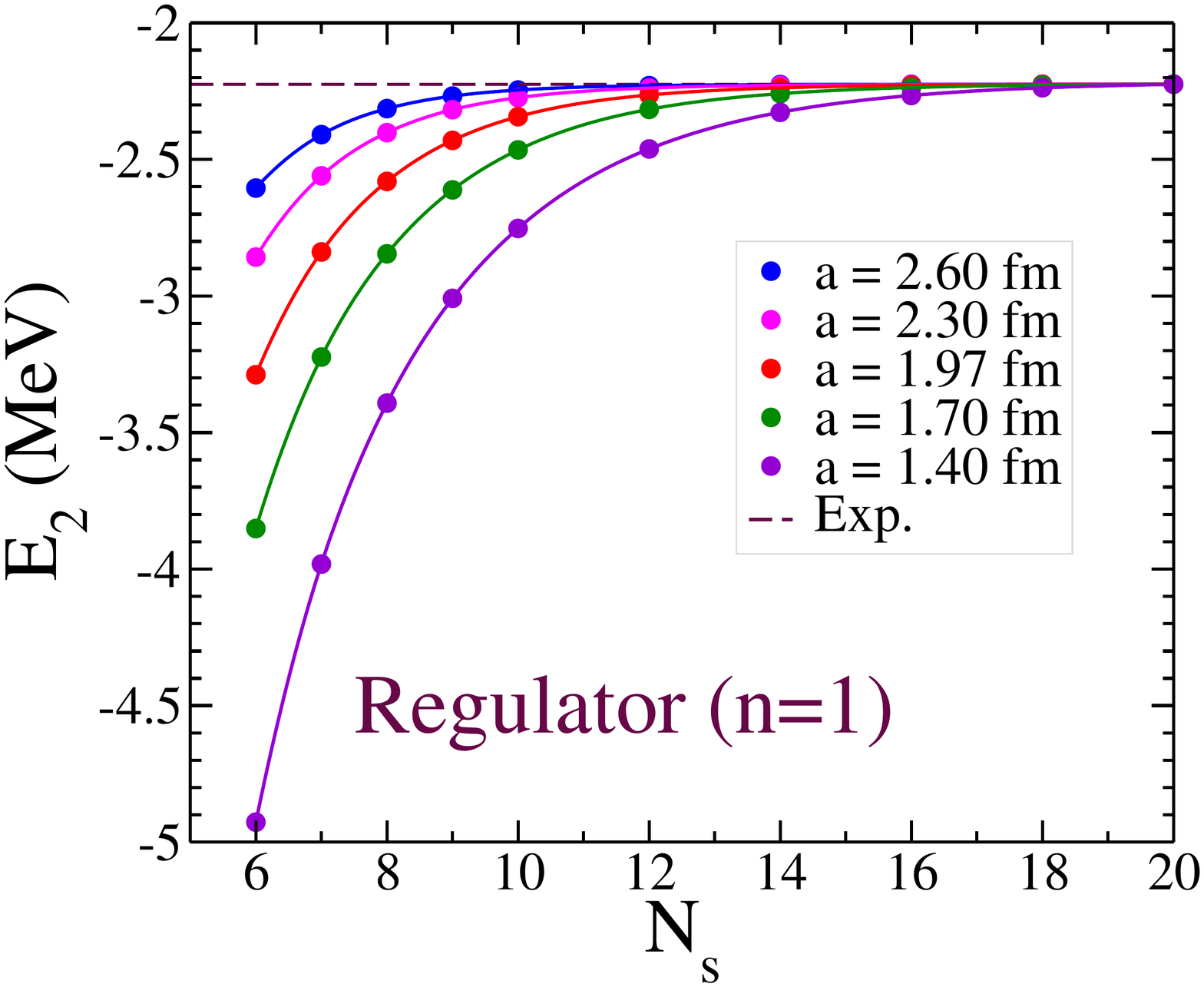} }}%
    \caption{Deuteron binding energy as a function of $N_s$ for different lattice spacing parameter $a$. In the left panel, the results are obtained with no regulator, whereas in the right panel, a regulator with power $n=1$ and the regulator parameter $b=0.01$ is applied.}%
    \label{Fig.BE_Ns}%
\end{figure*}

\begin{figure*}[htp!]
    \centering
    \subfloat{{\includegraphics[width=0.45\textwidth]{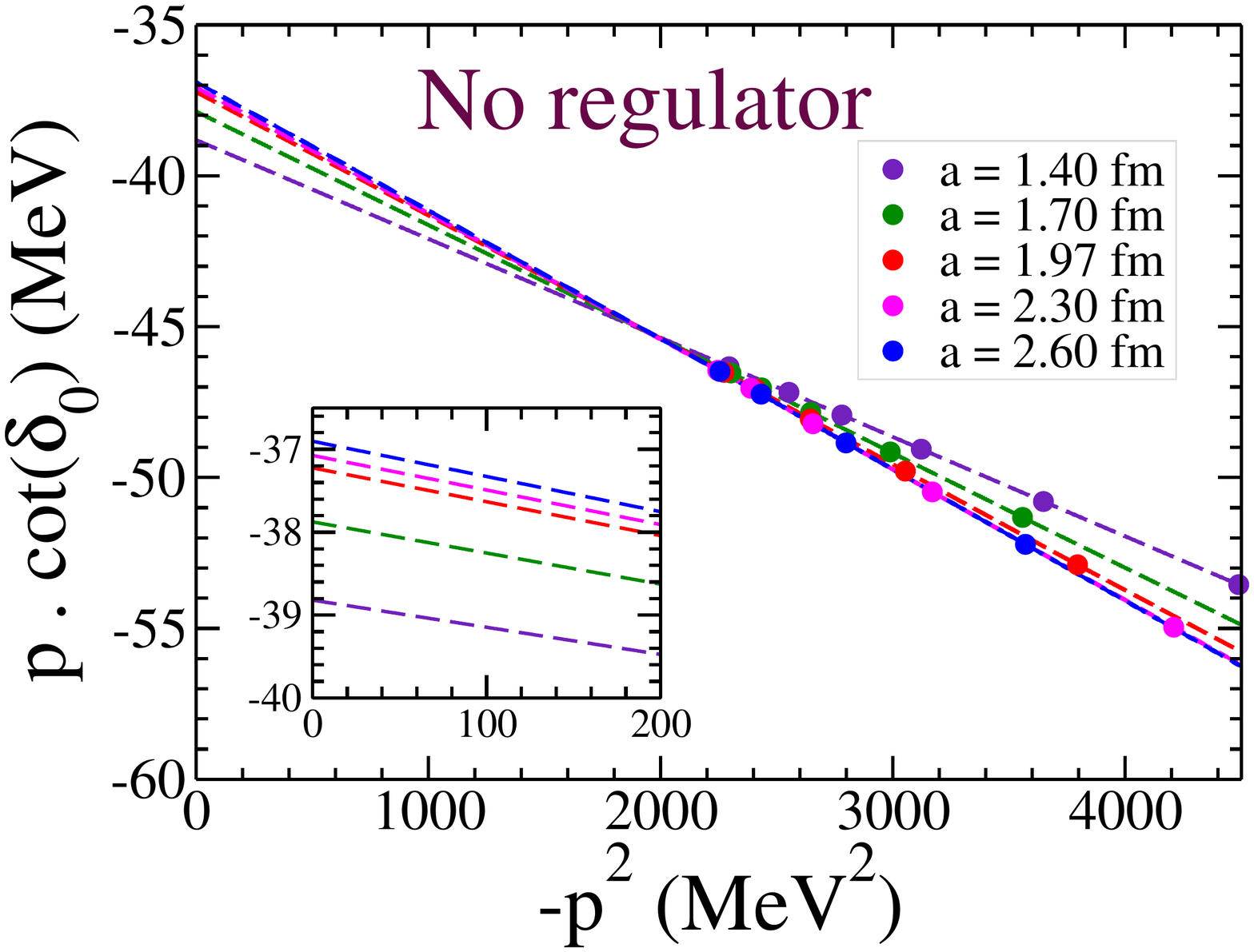} }}
    \subfloat{{\includegraphics[width=0.45\textwidth]{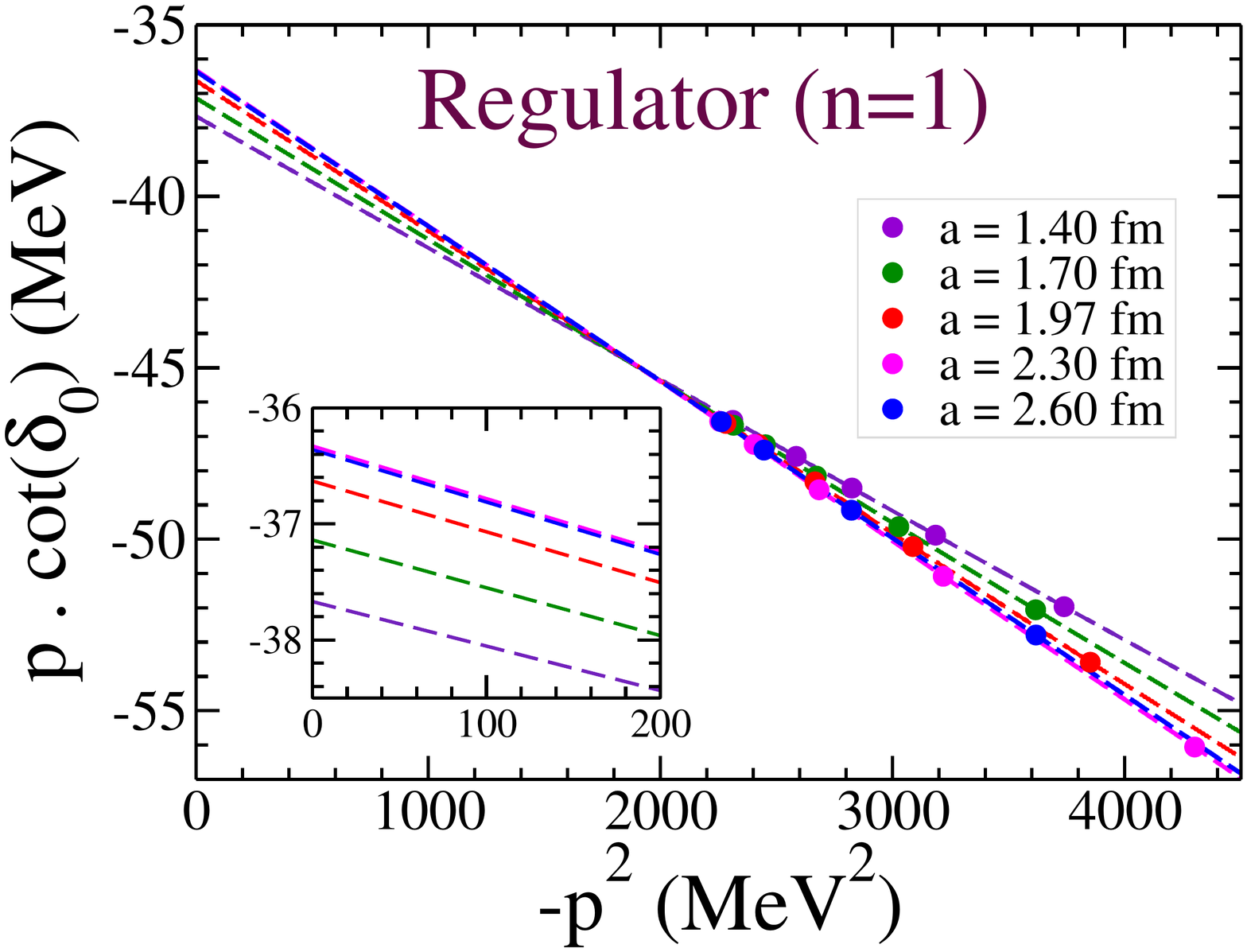} }}
    \caption{Effective range function in the $^3S_1$ neutron-proton channel for different values of lattice spacing parameter $a$.}%
    \label{fig_ERE_aset}%
\end{figure*}

In Fig. \ref{fig_ERE_1.97}, we have shown the effective range function, in the $^3S_1$ neutron-proton channel, calculated for lattice spacing $a=1.97$ fm as a function of the square of relative momentum. The results are shown at the LO and NLO.
As we have discussed earlier, by using a linear fit to our data and matching to Eq. (\ref{ERE}), one can extract the infinite volume ERE parameters from the finite volume energy eigenvalues.
The impact of different power of regulators (for $n=1, 2, 3$) on our data for the effective range function is shown.
As we can see, all regulators, independent of their power, are increasing the slope and decreasing the absolute value of the vertical intercept of the effective range function, indicating an increase in the scattering length and effective range parameter.
In the following, we discuss the impact of the regulator function on the ERE parameters extracted from different lattice spacing.
In the first step, we have calculated the lattice energy eigenvalues with and without the regularized interactions for different lattice spacing values. To this aim, we have considered a regulator with a power one. Starting with $N_s=20$, we extract the LECs $\Cz$ and $\Cd$ for different lattice spacing parameters $a=1.4, 1.7, 1.97, 2.3, 2.6$ fm, by fitting to the physical deuteron binding energy and ANC. This procedure leads to negative $\Cz$ and positive $\Cd$ for all considered lattice spacing parameters. Then by having the physical LECs, we have obtained a spectrum of the energy eigenvalues by lowering the number of nodes to $N_s=4$. Finally, by using L\"uscher formula in Eq. (\ref{ERE}), we extract the ERE parameters.
In Fig. \ref{Fig.BE_Ns}, our numerical results for deuteron binding energies obtained from the solution of Eq. (\ref{LS_lattice}), are shown as a function of the number of lattice nodes $N_s$, with and without using the regularized interactions. All the calculated energy eigenvalues used in Fig. \ref{Fig.BE_Ns} are given in the Appendix \ref{Appendix_Energies}.
The obtained effective range functions with different lattice spacing $a=1.4, 1.7, 1.97,2.3, 2.6$ fm are shown in Fig. \ref{fig_ERE_aset}. Our numerical results for extracted ERE parameters, with and without applying the regularization scheme, are presented in Table \ref{Table.overall_data}.
It should be noticed that the LECs $\Cz$ and $\Cd$ are fitted to the experimental values of deuteron binding energy and ANC with $N_s=20$. As we can see, the regularization scheme for lattice spacing greater than $2$ fm, brings the scattering length parameters $\at$ very close to the experimental value. Similarly,  the regularization scheme increases the effective ranges $\rt$ to values closer to the corresponding experimental value. 
So, we are confident that the introduced regularization scheme improves the extracted ERE parameters for different lattice spacing at NLO pionless EFT.
It should be mentioned that we have not manipulated the regularization parameter $b$ to reach the same ERE parameters for different lattice spacing. As it is shown earlier, the regulator parameter $b$ is dependent on the lattice spacing $a$ as $b = {\cal A}/ a^3$, while the value of ${\cal A}$ is considered to be constant for all lattice spacing. While the regularization scheme for smaller lattice spacing doesn't match the ERE parameters precisely to the corresponding experimental data, it brings them closer to the experimental data.
\begin{table}[htp!]
\caption{Deuteron binding energy, ANC and the ERE parameters $\at$ and $\rt$ calculated for different lattice spacing parameter $a$ with and without implementing the regularization scheme, suggested in Eqs. (\ref{regularization}) and (\ref{regulator}). The numbers in parentheses are the uncertainties in the last digits.}
\centering
\begin{ruledtabular}
\begin{tabular}{ccccccccccccc}
$ a$ (fm)  & $ \Cz$ & $ \Cd  \cdot 10^{-2}$ & $ E_2$ (MeV) & \ANC~(fm$ ^{-0.5}$) & $\at$ (fm) & $\rt$ (fm)   \\
\hline
& \multicolumn{6}{c}{{\bf No Regulator}}\\ \cline{2-7}
$ 1.4$  & $-2.142950$ & $+4.7378272$ & $-2.224575$ &   $ 0.249422$  &    $5.08(1)$ & $1.288(7)$  \\
$ 1.7$  & $-1.816890$ & $+3.7820497$ & $-2.224575$ &   $ 0.249424$  & $5.23(3)$ & $1.49(2)$  \\
$ 1.97$  & $-1.579070$ & $+3.0453214$ & $-2.224575$ &   $ 0.249425$  & $5.35(3)$ & $1.65(2)$  \\
$ 2.3$ & $-1.233188$ & $+1.9176492$ & $-2.224575$ &   $ 0.249423$  & $5.32(5)$ & $1.68(4)$  \\
$ 2.6$ & $-0.978510$ & $+1.2371360$ & $-2.224576$ &   $ 0.249426$  & $5.35(5)$ & $1.70(5)$  \\
\hline
& \multicolumn{6}{c}{{\bf With Regulator} ($ n=1, b = {\cal A}/a^3;  {\cal A}  =  0.07645373$ fm$^3$)}\\ \cline{2-7}
$ 1.4$ & $-2.064$ & $+5.1448251$ & $-2.224577$ &   $ 0.246912$  & $5.22(4)$ & $1.47(2)$    \\
$ 1.7$ & $-1.79465$ & $+3.9924816$ & $-2.224575$ &   $ 0.249424$  & $5.33(6)$ & $1.60(4)$  \\
$ 1.97$  & $-1.59677$ & $+3.2940765$ & $-2.224575$ & $ 0.249424$ & $5.43(5)$ & $1.74(4)$   \\
$ 2.3$ & $-1.323997$ & $+2.3189424$ & $-2.224575$ &   $ 0.249425$  & $5.43(3)$ & $1.81(3)$  \\
$ 2.6$ & $-1.02118$ & $+1.3854333$ & $-2.224574$ & $ 0.249427$ & $5.42(5)$ & $1.79(4)$   \\
\hline
Experiment & $-$ & $-$ & $-2.224575$ & $ 0.249424$ & $ 5.424(4)$ & $ 1.759(5)$ \\
\end{tabular}
\end{ruledtabular}
\label{Table.overall_data}
\end{table}

In Table \ref{Table.Comparison}, we have compared our ERE parameters extracted for lattice spacing $a=1.97$ fm, by different powers of the regulator, with the results of other studies.

\begin{table}[htp!]
\caption{Comparison of our ERE parameters in the $^3S_1$ channel, obtained with and without the application of the regularization scheme, with the results of other groups. The parameters ($n,b$) indicate the regulator parameters, introduced in Eq. (\ref{regulator}). The numbers in parentheses are the uncertainties in the last digits.}
\centering
\begin{ruledtabular}
\begin{tabular}{lllllllllllllllll}
Method & $ a$ & $\at$ (fm) & $\rt$ (fm) \\
\hline
Present ($n,b$)\\
 LO  $(1, 0)$ & $1.97$ fm & $4.577(7)$ & $0.496(8)$   \\
 LO  $(1, 0.01)$ & $1.97$ fm &  $4.652(7)$ & $0.621(7)$   \\
 LO  $(2, 0.01)$ & $1.97$ fm &  $4.624(7)$ & $0.580(8)$    \\
 LO  $(3, 0.01)$ & $1.97$ fm &  $4.64(1)$  & $0.60(1)$   \\
 NLO  $(1, 0)$ & $1.97$ fm &  $5.35(3)$ & $1.65(2)$    \\
  NLO  $(1, 0.01)$ & $1.97$ fm &   $5.43(5)$ & $1.74(4)$   \\
  NLO  $(2, 0.01)$ & $1.97$ fm &   $5.41(4)$ & $1.73(3)$    \\
 NLO  $(3, 0.01)$ & $1.97$ fm &   $5.42(3)$ & $1.76(2)$   \\
\hline
Borasoy {\it et al.} (LO Pionless EFT) \cite{borasoy2007lattice} & $1.97$ fm  & $ 4.522(1)$ & $ 0.30(2)$ \\
 & $1.97$ fm  & $ 4.664(1)$ & $ 0.53(2)$ \\
 \hline
 Rokash {\it et al.} (LO Pionless EFT) \cite{rokash2013finite}& $ 2$  fm & $ 4.50$ & $ 0.33$ \\
 \hline
Klein {\it et al.}  (LO Pionless EFT) \cite{klein2015regularization} & $1.97$ fm  & $ 5.611(1)$ & $ 2.029(1)$ \\
Klein {\it et al.} (LO pionfull EFT) \cite{klein2015regularization} & $1.97$ fm  & $ 5.470(1)$ & $ 1.818(1)$ \\
\hline
Alarc{\'o}n {\it et al.} (LO pionfull EFT) \cite{alarcon2017neutron}  & $1.97$ fm & $5.46(1)$ & $1.686(1)$ \\
Alarc{\'o}n {\it et al.} (NLO pionfull EFT) \cite{alarcon2017neutron}  & $1.97$ fm & $5.31(2)$ & $1.79(3)$ \\
Alarc{\'o}n {\it et al.} (N$^2$LO pionfull EFT) \cite{alarcon2017neutron}  & $1.97$ fm & $5.35(2)$ & $1.82(3)$ \\
\hline
Experiment & & $ 5.424(4)$ & $ 1.759(5)$ \\
\end{tabular}
\end{ruledtabular}
\label{Table.Comparison}
\end{table}

\section{Conclusion}\label{conclusion}

In this paper, we have studied the impact of a new regularization scheme on the extraction of the ERE parameters of $^3S_1$ channel for different lattice spacing in a pionless effective field theory up to NLO. 
We first use the deuteron binding energy and the ANC to fix the LECs of the contact interactions by solving the lattice form of the Lippmann-Schwinger equation with Lanczos technique.
Then we employ L\"uscher's finite-volume relation to extract the $S-$wave ERE parameters $\rt$ and $\at$ from the lattice energy eigenvalues corresponding to the different lattice size.
The lattice spacing dependence of the ERE parameters is studied in the range $1.4 \le a \le 2.6$ fm.
To eliminate the lattice artifacts, an ${\cal O} (a^4)$-improved lattice action is considered.
The impact of different powers of the exponential regulator is studied for the lattice spacing $a=1.97$ fm, and it is shown that they have an almost similar influence on the extracted ERE parameters. 
The introduced regulator is applied to different lattice spacing, leading to an improvement on the extraction of the ERE parameters, and brings them close to the experimental data for $a \ge 2$ fm.


\begin{acknowledgments}
We thank Koji Harada for sharing their results, which allowed us to validate our codes for the solution of the Lippmann-Schwinger equation for two-nucleon-bound states on a lattice.
The work of M. R. Hadizadeh was supported by the National Science Foundation under grant NSF-PHY-2000029 with Central State University.

\end{acknowledgments}

\appendix

\section{Two-Nucleon Energy Eigenvalues}\label{Appendix_Energies}

In Tables \ref{Table_a1.4}-\ref{Table_a2.6}, we provide our numerical results for the solution of the Lippmann-Schwinger equation, given in Eq. (\ref{LS_lattice}), with the LECs given in Table \ref{Table.overall_data2}, for different values of lattice spacing parameter $a$ and different number of lattice nodes $N_s$.

\begin{table*}[htp!]
\caption{The LECs $\Cz$ and $\Cd$ fitted to deuteron binding energy and ANC for different lattice spacing parameter $a$ with and without implementing the regularization scheme, introduced in Eqs. (\ref{regularization}) and (\ref{regulator}). $n$ indicates the power of the exponential regulator.}
\centering
\begin{ruledtabular}
\begin{tabular}{ccccccccccccc}
$ a$ (fm)  & $ \Cz$ & $ \Cd  \cdot 10^{-2}$    \\
\hline
& \multicolumn{2}{c}{{\bf No Regulator}}\\ \cline{2-3}
$ 1.4$  & $-2.142950$ & $+4.7378272$   \\
$ 1.7$  & $-1.816890$ & $+3.7820497$  \\
$ 1.97$  & $-1.579070$ & $+3.0453214$   \\
$ 2.3$ & $-1.233188$ & $+1.9176492$   \\
$ 2.6$ & $-0.978510$ & $+1.2371360$  \\
\hline
& \multicolumn{2}{c}{{\bf With Regulator} ($ n=1$)}\\ \cline{2-3}
$ 1.4$ & $-2.064$ & $+5.1448251$     \\
$ 1.7$ & $-1.79465$ & $+3.9924816$   \\
$ 1.97$  & $-1.59677$ & $+3.2940765$    \\
$ 2.3$ & $-1.323997$ & $+2.3189424$  \\
$ 2.6$ & $-1.02118$ & $+1.3854333$    \\
\hline
& \multicolumn{2}{c}{{\bf With Regulator} ($ n=2$)}\\ \cline{2-3}
$ 1.97$  & $-1.587607$ & $+3.4237993$    \\
\hline
& \multicolumn{2}{c}{{\bf With Regulator} ($ n=3$)}\\ \cline{2-3}
$ 1.97$  & $-1.574284$ & $+3.945544$    \\
\end{tabular}
\end{ruledtabular}
\label{Table.overall_data2}
\end{table*}

\begin{table*}[htp!]
\caption{Deuteron binding energy calculated for different values of $N_s$ with the lattice spacing $a= 1.4$ fm. The parameters ($n,b$) indicate the regulator parameters, introduced in Eq. (\ref{regulator}).}
\centering
\begin{ruledtabular}
\begin{tabular}{ccccccccccccc}
$ N_s$  & NLO ($ n=1, b=0$)   & NLO ($ n=1, b=2.786215 \cdot 10^{-2}$)   \\
\hline
$ 20$ & $-2.224575$ & $-2.224577$ \\
$ 18$ & $-2.236165$ & $-2.237620$ \\
$ 16$ & $-2.261734$ & $-2.265850$ \\
$ 14$ & $-2.318403$ & $-2.327441$ \\
$ 12$ & $-2.443624$ & $-2.461672$ \\
$ 10$ & $-2.717962$ & $-2.752608$ \\
$ 9$ & $-2.960845$ & $-3.008798$ \\
$ 8$ & $-3.325327$ & $-3.392493$ \\
$ 7$ & $-3.885172$ & $-3.981742$ \\
$ 6$ & $-4.781351$ & $-4.926780$ \\
$ 5$ & $-6.316952$ & $-6.553276$ \\
$ 4$ & $-9.250943$ & $-9.680334$ \\
\end{tabular}
\end{ruledtabular}
\label{Table_a1.4}
\end{table*}

\begin{table*}[htp!]
\caption{The same as Table \ref{Table_a1.4}, but for $a=1.7$ fm.}
\centering
\begin{ruledtabular}
\begin{tabular}{ccccccccccccc}
$ N_s$  & NLO ($ n=1, b=0$)   & NLO ($ n=1, b= 1.556152 \cdot 10^{-2}$)   \\
\hline
$ 20$ & $-2.224575$ & $-2.224577$ \\
$ 18$ & $-2.227704$ & $-2.228103$ \\
$ 16$ & $-2.235686$ & $-2.236853$ \\
$ 14$ & $-2.256377$ & $-2.259115$ \\
$ 12$ & $-2.310365$ & $-2.316613$ \\
$ 10$ & $-2.452207$ & $-2.465131$ \\
$ 9$ & $-2.592639$ & $-2.611437$ \\
$ 8$ & $-2.818116$ & $-2.845511$ \\
$ 7$ & $-3.183195$ & $-3.223659$ \\
$ 6$ & $-3.789626$ & $-3.851409$ \\
$ 5$ & $-4.852515$ & $-4.953149$ \\
$ 4$ & $-6.904684$ & $-7.086130$ \\
\end{tabular}
\end{ruledtabular}
\label{Table_a1.7}
\end{table*}

\begin{table*}[htp!]
\caption{The same as Table \ref{Table_a1.4}, but for $a=1.97$ fm.}
\centering
\begin{ruledtabular}
\begin{tabular}{ccccccccccccc}
$ N_s$  & \multicolumn{4}{c}{{\bf LO }}  & \multicolumn{4}{c}{{\bf NLO }}  \\ \cline{2-5} \cline{7-10}
 & No Reg. & $ n=1$ & $ n=2$ & $ n=3$  && No Reg. & $ n=1$  & $ n=2$ & $ n=3$ \\
\hline
$ 20$ & $-2.224575$ & $-2.224573$ & $-2.224576$ & $-2.224575$  && $-2.224575$ & $-2.224575$ & $-2.224574$ & $-2.224575$ \\
$ 18$ & $-2.225251$ & $-2.225346$ & $-2.225263$ & $-2.225265$  && $-2.225550$ & $-2.225704$ & $-2.225583$ & $-2.225610$ \\
$ 16$ & $-2.227186$ & $-2.227485$ & $-2.227247$ & $-2.227244$  && $-2.228361$ & $-2.228816$ & $-2.228490$ & $-2.228589$ \\
$ 14$ & $-2.232922$ & $-2.233633$ & $-2.233106$ & $-2.233127$  && $-2.236632$ & $-2.237740$ & $-2.237041$ & $-2.237354$ \\
$ 12$ & $-2.250310$ & $-2.251998$ & $-2.250869$ & $-2.251025$  && $-2.261507$ & $-2.264124$ & $-2.262728$ & $-2.263680$ \\
$ 10$ & $-2.304263$ & $-2.308227$ & $-2.305968$ & $-2.306441$  && $-2.337196$ & $-2.343432$ & $-2.340805$ & $-2.343569$ \\
$ 9$ & $-2.365196$ & $-2.371330$ & $-2.368068$ & $-2.368943$  && $-2.420902$ & $-2.430523$ & $-2.426944$ & $-2.431602$ \\
$ 8$ & $-2.472692$ & $-2.482165$ & $-2.477541$ & $-2.479011$  && $-2.565747$ & $-2.580570$ & $-2.575767$ & $-2.583490$ \\
$ 7$ & $-2.661788$ & $-2.676394$ & $-2.669839$ & $-2.672291$  && $-2.815834$ & $-2.838765$ & $-2.832374$ & $-2.845091$ \\
$ 6$ & $-2.995661$ & $-3.018378$ & $-3.009063$ & $-3.013079$  && $-3.252139$ & $-3.288325$ & $-3.279835$ & $-3.301075$ \\
$ 5$ & $-3.598996$ & $-3.635343$ & $-3.621888$ & $-3.628761$  && $-4.041911$ & $-4.102015$ & $-4.091016$ & $-4.129732$ \\
$ 4$ & $-4.755451$ & $-4.817499$ & $-4.797340$ & $-4.811362$  && $-5.592160$ & $-5.701678$ & $-5.683439$ & $-5.743210$ \\
\end{tabular}
\end{ruledtabular}
\label{Table_a1.97}
\end{table*}

\begin{table*}[htp!]
\caption{The same as Table \ref{Table_a1.4}, but for $a=2.3$ fm.}
\centering
\begin{ruledtabular}
\begin{tabular}{ccccccccccccc}
$ N_s$  & NLO ($ n=1, b=0$)   & NLO ($ n=1, b=6.283696 \cdot 10^{-3}$)   \\
\hline
$ 20$ & $-2.224575$ & $-2.224575$ \\
$ 18$ & $-2.224804$ & $-2.224863$ \\
$ 16$ & $-2.225572$ & $-2.225748$ \\
$ 14$ & $-2.228180$ & $-2.228647$ \\
$ 12$ & $-2.237352$ & $-2.238583$ \\
$ 10$ & $-2.270483$ & $-2.273990$ \\
$ 9$ & $-2.312144$ & $-2.318053$ \\
$ 8$ & $-2.391733$ & $-2.401826$ \\
$ 7$ & $-2.542779$ & $-2.559982$ \\
$ 6$ & $-2.828306$ & $-2.857706$ \\
$ 5$ & $-3.375589$ & $-3.427266$ \\
$ 4$ & $-4.483011$ & $-4.581039$ \\
\end{tabular}
\end{ruledtabular}
\label{Table_a2.3}
\end{table*}

\begin{table*}[htp!]
\caption{The same as Table \ref{Table_a1.4}, but for $a=2.6$ fm.}
\centering
\begin{ruledtabular}
\begin{tabular}{ccccccccccccc}
$ N_s$  & NLO ($ n=1, b=0$)   & NLO ($ n=1, b=4.3498936 \cdot 10^{-3}$)   \\
\hline
$ 20$ & $-2.224576$ & $-2.224574$   \\
$ 18$ & $-2.224638$ & $-2.224659$   \\
$ 16$ & $-2.224869$ & $-2.224938$   \\
$ 14$ & $-2.225771$ & $-2.225941$   \\
$ 12$ & $-2.229746$ & $-2.229861$   \\
$ 10$ & $-2.244550$ & $-2.245892$   \\
$ 9$   & $-2.265714$ & $-2.268085$   \\
$ 8$   & $-2.309831$ & $-2.314116$   \\
$ 7$   & $-2.401551$ & $-2.409350$   \\
$ 6$   & $-2.590731$ & $-2.604906$   \\
$ 5$   & $-2.979936$ & $-3.005941$   \\
$ 4$   & $-3.802672$ & $-3.852706$   \\
\end{tabular}
\end{ruledtabular}
\label{Table_a2.6}
\end{table*}

\bibliography{Refs}

\end{document}